\documentclass[aps,showkeys]{revtex4}
\usepackage{amssymb}
\usepackage{epsf}    
\usepackage{wrapfig}
\usepackage{amssymb}
 
\def\sgg{g_{\sigma\gamma\gamma}}
\def\gp{g_{\sigma\pi\pi}}
\def\ggg{\Gamma_{\sigma\gamma\gamma}}
\def\am{(\alpha_1-\beta_1)}
\def\ap{(\alpha_1+\beta_1)}

\def\comp{\gamma+\pi^{\pm}\to\gamma+\pi^{\pm}}
\def\sigp{d\sigma_{\gamma\pi\to\gamma\pi}}

\def\sip{\frac{d\sigma_{\gamma\pi}}{d\Omega}}
\def\si0{\frac{d\sigma^0_{\gamma\pi}}{d\Omega}}
\def\thq{\theta_{\gamma\gamma}^{cm}}

\begin{document}

\title{\bf{DIPOLE POLE POLARIZABILITIES OF  $\pi^{\pm}$--MESONS}}

\author{L.V.~Fil'kov$^1$\footnote[1]{filkov@sci.lebedev.ru} and
V.L.~Kashevarov$^{2,1}$  
\vspace*{0.3cm}}

\affiliation{$^1$Lebedev Physical Institute, 119991 Moscow, Russia\\
$^2$Institut f\"ur Kernphysik, Johannes Gutenberg-Universit\"at Mainz,
55099 Mainz, Germany}

\begin{abstract}
The main experimental works, where dipole polarizabilities of
charged pions have been determined, are  considered.
Possible reasons for the differences between the experimental data are 
discussed. In particular, it is shown that the account of the
$\sigma$-meson gives a significant correction to the value of the 
polarizability obtained in the latest experiment of the COMPASS 
collaboration. We present also new fit results for the 
$(\gamma\gamma\to\pi\pi)$ reaction.

\end{abstract}


\maketitle                                                                               

\section{Introduction}

Pion polarizabilities are fundamental structure parameters 
values of which are very sensitive to predictions of
different theoretical models. Therefore, an accurate experimental determination 
of these parameters is very important for testing the validity of such models.

The most of experimental data obtained for the difference of the electric 
$(\alpha_1)$ and $(\beta_1)$ dipole 
polarizabilities of the charged pions are presented in Table I. 

The polarizabilities were determined by analyzing the processes of the high 
energy pions scattering in the Coulomb field of heavy nuclei 
$(\pi^-A\to\gamma\pi^-A')$ via the Primakoff effect,
radiative pion photoproduction from proton $(\gamma p\to\gamma\pi^+n)$, and 
two-photon production of pion pairs $(\gamma\gamma\to\pi\pi)$.
As seen from Table I, the data vary from 4 up to 40 and are in conflict even 
for experiments performed with the same method.
In this paper we will consider possible reasons for such disagreements.

\begin{table*}
\caption{Review of experimental data on $\am_{\pi^{\pm}}$} 
\centering
\begin{tabular}{|ll|l|} \hline
\multicolumn{2}{|c|}{Experiments} & \qquad\quad $\am_{\pi^{\pm}}$ \\ \hline
$\gamma p\to\gamma\pi^+n$ & MAMI (2005) \cite{mami} 
& $11.6\pm 1.5_{stat}\pm 3.0_{syst}\pm 0.5_{mod}$ \\ \hline
$\gamma p\to\gamma\pi^+n$ & Lebedev Phys. Inst. (1984) \cite{lebed}& 
$40\pm 20$ \\ \hline
$\pi^-A\to\gamma\pi^-A'$ & Serpukhov (1983) \cite{serp} & $13.6\pm 2.8\pm 2.4$ 
\\ \hline  
$\pi^-A\to\gamma\pi^-A'$ & COMPASS (2007) \cite{comp} & $4.0\pm 1.2\pm 1.4 $ 
\\ \hline

\multicolumn{2}{|c|}
{$\gamma\gamma\to\pi^+\pi^-$, D. Babusci {\em et al.} (1992) \cite{bab}} &   \\
     & PLUTO \cite{pluto}   & $38.2\pm 9.6\pm 11.4$ \\
     & DM 1 \cite{dm1}     & $34.4\pm 9.2$  \\
     & MARK II \cite{mark} & $4.4\pm 3.2$    \\ \hline
\multicolumn{2}{|c|}
{J.F. Donoghue, B.R. Holstein (1993) \cite{don}, Mark II \cite{mark}} &  5.4 \\
\hline
\multicolumn{2}{|c|}  
{A.E. Kaloshin, V.V. Serebryakov (1994) \cite{kal}, Mark II \cite{mark}} & 
$5.25\pm 0.95$ \\ \hline
\multicolumn{2}{|c|}
{L.V. Fil'kov, V.L. Kashevarov (2006) \cite{fil3}}&  $13^{+2.6}_{-1.9}$ \\
&the fit to the data \cite{mark,tpc,cello,ven,aleph,belle} up to 
2.5 GeV & \\ \hline
\multicolumn{2}{|c|}
{R. Garcia-Martin, B. Moussallam (2010) \cite{garcia}}  & 4.7 \\  \hline
 
\end{tabular}
\end{table*}
                                                                                      
\section{Scattering of pions in the Coulomb field of heavy nuclei}

The charged pion polarizability was obtained at the first time 
in the work \cite{serp} from the scattering of $\pi^-$ mesons 
off the Coulomb field of heavy nuclei. 
                                                              
The cross section of the radiative pion scattering $\pi A\to\pi\gamma A'$  
via the Primakoff effect can be written as
\begin{equation}
\label{cross}
\frac{d\sigma_{\pi A}}{ds dQ^2 d \cos\thq} 
=\frac{Z^2 \alpha}{\pi (s-\mu^2)} 
 F^2_{eff}(Q^2)\frac{Q^2-Q^2_{min}}{Q^4} \frac{d\sigma_{\pi\gamma}}{d\cos\thq},
\end{equation} 
where $F_{eff}\approx 1$ is the electromagnetic form-factor of nucleus,
$\alpha$ is the fine-structure constant, $Z$ is the charge number of 
the nucleus,
and $Q^2$ is the negative 4-momenta transfer squared, $Q^2=-(p_A-p_A')^2$.
$Q^2_{min}$ is the minimum value of $Q^2$ which is given by the formula
$Q^2_{min}=(s-\mu^2)^2/(4 E^2_{beam})$,
where $s$ is the square of the total energy of the process $\comp$,
$E_{beam}$ is the pion beam energy.  
This cross section has a Coulomb peak at $Q^2=2 Q^2_{min}$ with a width equal 
to $\simeq 6.8 Q^2_{min}$.

The experiment \cite{serp} was carried at a beam energy equal to 40 GeV.
In this case if the energy of the incident photon in the pion rest 
frame $\omega_1=600$ MeV, then $Q^2_{min}$ is equal to $4.4\times 10^{-6}$
(GeV/c)$^2$. It was shown that the Coulomb amplitude dominates in this case 
for $Q^2 \leq 2\times 10^{-4}$(GeV/c)$^2$. The experiment \cite{serp} was 
carried out at $Q^2_{cut}<6\times 10^{-4}$(GeV/c)$^2$. Events in the region of 
$Q^2$ of $(2-8)\times 10^{-3}$(GeV/c)$^2$ were used to estimate the 
strong interaction background. This background was assumed to behave either
as $\sim Q^2$ in the region $Q^2 \leq 6\times 10^{-4}$(GeV/c)$^2$ or as
a constant. The polarizability was determined from the ratio (assuming
$\ap_{\pi^{\pm}}=0$)
\begin{equation}
\label{R}
R_{\pi}=(\sip)/(\si0)=1-\frac{3}{2}\frac{\mu^3}{\alpha}
\frac{x_{\gamma}^2}{1-x_{\gamma}}\alpha_{\pi},
\end{equation}
where $\sip$ refers to the measured cross section and $\si0$ to simulated 
cross section expected for $\alpha_{\pi}=0$, $x_{\gamma}=E_{\gamma}/E_{beam}$ 
in the laboratory system of the process $\pi A\to\pi\gamma A'$. 
As a result they have obtained: $\am_{\pi^{\pm}}=13.6 \pm 2.8 \pm 2.4$ .

The new result of the COMPASS collaboration \cite{comp}
($\alpha_{\pi}=2.0\pm 0.6_{stat.}\pm 0.7_{syst.}$) has 
been found also by studying the $\pi^-$-meson scattering off the Coulomb field 
of heavy nuclei. This value is at variance with the result  obtained in a 
very similar experiment in Serpukhov \cite{serp}, but also with \cite{mami}. 
 
This experiment \cite{comp} was performed with $E_{beam}=190$ GeV.  
For such values of $E_{beam}$ the quantity of $Q^2_{min}(COMPASS)$ must be 
smaller by 22.5 times than $Q^2_{min}$(Serpukhov). 
However, the authors of the experiment \cite{comp} considered 
$Q^2_{cut}\lesssim 0.0015$ (GeV/c)$^2$, which are essentially greater 
than $Q^2_{cut}$ in work \cite{serp}.
 
As shown in \cite{faldt1} the basic ratio $R_{\pi}$ is applicable for the 
Coulomb peak only. On the other hand, in Ref.~\cite{faldt2} it is 
elaborated that the Coulomb 
amplitude interference with the coherent nuclear amplitude is important for 
$0.0005 \leqslant Q^2\leqslant 0.0015$\,(GeV/c)$^2$. 
This means that the Serpukhov analysis could safely apply the ratio $R_{\pi}$ 
in (\ref {R}), whereas COMPASS has to consider the interference of the Coulomb 
and strong amplitudes. The phase determined with the simple considerations in 
Ref.~\cite{walch} for the Serpukhov experiment \cite{serp} is close to $\pi/2$ 
meaning that the subtraction of a nuclear background assumed to be incoherent 
is justified. 
 
In Refs. \cite{faldt1,faldt2}, the strong amplitude is described by the
Glauber model (elastic multiple scattering of hadrons in nuclei). 
The conditions and limitations of the Glauber approximation are discussed 
in the classical article about diffraction by U. Amaldi, M. Jacob, 
and G. Matthiae \cite{amaldi}. G\"{o}ran F\"{a}ldt and Ulla Tengblad  
\cite{faldt1,faldt2}  assume that the hadron-nucleon potential of
the nucleons in the nucleus is local and also real, then the phases between the 
incoming hadron and the nucleons add up linearly. However, at high energies -  
and the COMPASS energy of the incoming $\pi$ with 180\,(GeV/c)$^2$ is high - the 
strong phases become complex and the summed amplitude acquires an additional 
energy dependent phase.  The associated profile function must take into account 
multiple scattering and will be complex, i.e. an unknown phase appears.  

Moreover, the simulation of the distribution in Fig. 3(c) in the  work 
(\cite{comp}) does not reproduce the diffraction bumps at $Q>0.04$\,(GeV/c).
With a more realistic 
"absorbing disc" for the profile function \cite{amaldi} all 
bumps in Fig.~3(c) could be reproduced well and again a phase would be close 
to $\pi/2$ \cite{WalchPrivat}.
Without a real fit to the data it is impossible to estimate the 
effect of the model dependence of the diffractive background, but that it will 
have an influence is clear from Ref.~\,\cite{faldt2}. 

Comparison of data with different targets provides the possibility to check
the $Z^2$ dependence for the Primakoff cross section and estimate a possible
contribution of the nuclear background. Such an investigation was performed
by the Serpukhov collaboration and they have obtained $Z^2$ dependence with
good enough accuracy. The COMPASS collaboration really have gotten their main
result using only $Ni$ target but they wrote that they also considered other 
targets on small statistic and obtained approximate $\sim Z^2$  dependence.

It should be noted that in order to get an information about the pion 
polarizabilities, the authors considered the cross section of the process 
$\gamma\pi^-\to\gamma\pi^-$ equal to the Born cross section and the interference 
of the Born amplitude with the pion polarizabilities only. The COMPASS 
collaboration analyzed this process up to the total energy $W=490$ MeV in the 
angular range $0.15>\cos\theta^{cm}_{\gamma\pi}>-1$. However, the contribution 
of the $\sigma$-meson to the cross section of the Compton scattering on the pion 
could be very substantial in this region of the energy and angles \cite{fil1}.  
Therefore, we consider this contribution.

\section{$\sigma$-meson contribution}

According the dispersion relation (DR) from Ref.~\cite{fil1} the contribution
of the $\sigma$-meson can be determined as 
\begin{equation}
Re M_{++}^{\sigma} = \frac{t}{\pi}\int\limits_{4\mu^2}^{\infty}~\frac{
Im M_{++}^{\sigma}(t',s=\mu^2)~d t'}{t'(t'-t)}.
\end{equation}

The imaginary amplitude $Im M_{++}^{\sigma}(t,s=\mu^2)$ has to be evaluated 
taking into account that the $\sigma$-meson is a pole on the second Riemann 
sheet. The relation between amplitudes on the first and the second sheets can 
be written \cite{oller} as
\begin{equation}
\label{f0}
F_0^{II}(t+i\epsilon)=F_0^{I}(t+i\epsilon)(1+2i\rho T_0^{II}(t+i\epsilon)),
\end{equation}     
where
\begin{equation}                                                          
T_0^{II}=-\frac{\gp^2}{t_{\sigma}-t}, \quad  
F_0^{II}=\sqrt{2}\,\frac{\sgg \gp}{t_{\sigma}-t},
\end{equation}
\begin{equation}
t_{\sigma}=(M_{\sigma}-i\Gamma_{\sigma 0}/2)^2, \quad  
\rho=\frac{\sqrt{1-4\mu^2/t}}{16\pi}, \quad
\Gamma_{\sigma 0}=\Gamma_{\sigma}\left( \frac{t-4\mu^2}
{M_{\sigma}^2-4\mu^2}\right)^{1/2}.
\end{equation}

Using the relation (\ref{f0}) we have
\begin{equation}
\label{imm}
Im M_{++}^{\sigma}(t,s=\mu^2)=\frac{1}{t}\sqrt{\frac{2}{3}}\,\frac{\sgg \gp R}
{D^2+R^2}, \;  D=(M_{\sigma}^2-t-\frac{1}{4}\Gamma_{\sigma 0}^2), \; 
R=M_{\sigma}\Gamma_{\sigma 0}+2\rho\gp^2.
\end{equation}

We can get influence of the $\sigma$-meson on the extracted value of 
$\am_{\pi^{\pm}}$
by equating the cross section  without $\sigma$-meson contribution to the
cross section when $\sigma$-meson is taking into account \cite{fil4}:
\begin{equation}
\label{rel}
\sigp(B, \am_{\pi^{\pm}}^0)/d\Omega 
=\sigp(B, M_{++}^{\sigma} \am_{\pi^{\pm}})/d\Omega,
\end{equation}
where
$\am_{\pi^{\pm}}^0$ is the value of $\am_{\pi^{\pm}}$ without of the 
$\sigma$ contribution obtained in \cite{comp} and $B$ is the Born term.

For backward scattering ($z=-1$), we have the following expression:
\begin{equation}
\label{amc1}
\am_{\pi^{\pm}}=\frac{1}{4\pi\mu}\left\{-(B+Re M_{++}^{\sigma}) 
     +\frac{B^2+4\pi\mu B\am_{\pi^{\pm}}^0}{B+Re M_{++}^{\sigma}}\right\},
 \; B=\frac{2e^2\mu^2}{(s-\mu^2)(u-\mu^2)}.
\end{equation}  

\begin{figure}
\epsfxsize=6cm      
\epsfysize=7cm      
\centerline{              
\epsffile{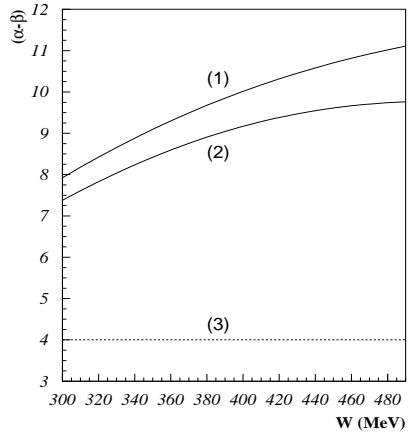}}              
\caption{Dependence of $\am_{\pi^{\pm}}$ on $W$. Lines (1) and (2)
correspond to the calculation of Eqs. (\ref{amc1}) and  (\ref{amc2}), respectively.
Line (3) is the COMPASS result \cite{comp}.}
\label{(a-b)1}
\end{figure}

In the case of integration over the region $-1\leq z\leq 0.15$ we have
\begin{equation}
\label{amc2}
\am_{\pi^{\pm}}=F_0/F_1,           
\end{equation}
where
\begin{equation}
\label{f01}           
F_0 =\frac{1}{4\pi\mu}\left\{\int\limits_{-1}^{0.15}~\left[-Re M_{++}^{\sigma}
  (Re M_{++}^{\sigma}+2 B) +4\pi\mu B\am_{\pi^{\pm}}^0\right](1-z)^2~dz \right\},
\end{equation} 
\begin{equation}
\label{f1}
F_1=\left\{\int\limits_{-1}^{0.15}~(B+Re M_{++}^{\sigma})(1-z)^2~dz\right\}.
\end{equation}     
In the calculation we used the parameters of the $\sigma$-meson from 
Ref.~\cite{oller}:
$M_{\sigma}=441$MeV, $\Gamma_{\sigma}=544$MeV, $\ggg=1.98$keV, $\gp=3.31$GeV, 
$\sgg^2=16\pi\ggg M_{\sigma}$.
The results of the calculations using Eq.~(\ref{amc1}) (line (1) ) 
and Eq.~(\ref{amc2}) (line (2)) are shown in Fig.~1. Line (3) is the result of 
Ref.~\cite{comp}. As a result we have obtain $\am_{\pi^{\pm}} \sim 10$. However 
the magnitude of $\am_{\pi^{\pm}}$ is very sensitive to parameters of the 
$\sigma$-meson and can reach a value of $\sim 11$ for the parameters from 
\cite{penn}.

So, the contribution of the $\sigma$-meson can essentially change the COMPASS 
result. It should be noted that the contribution of the 
$\sigma$-meson was not considered in Serpukhov as well. 
However, in this case, the contribution of the $\sigma$-meson for the Serpukhov
kinematics is $\Delta (\alpha-\beta)_{\sigma}\gtrsim 2.7$ within the
experimental error of the Serpukhov result.

\section{Two-photon production of pion pairs}
                                      
\begin{figure*}[t]\label{crosy}
\epsfxsize=11.5cm
\epsfysize=10.0cm
\centerline{
\epsffile{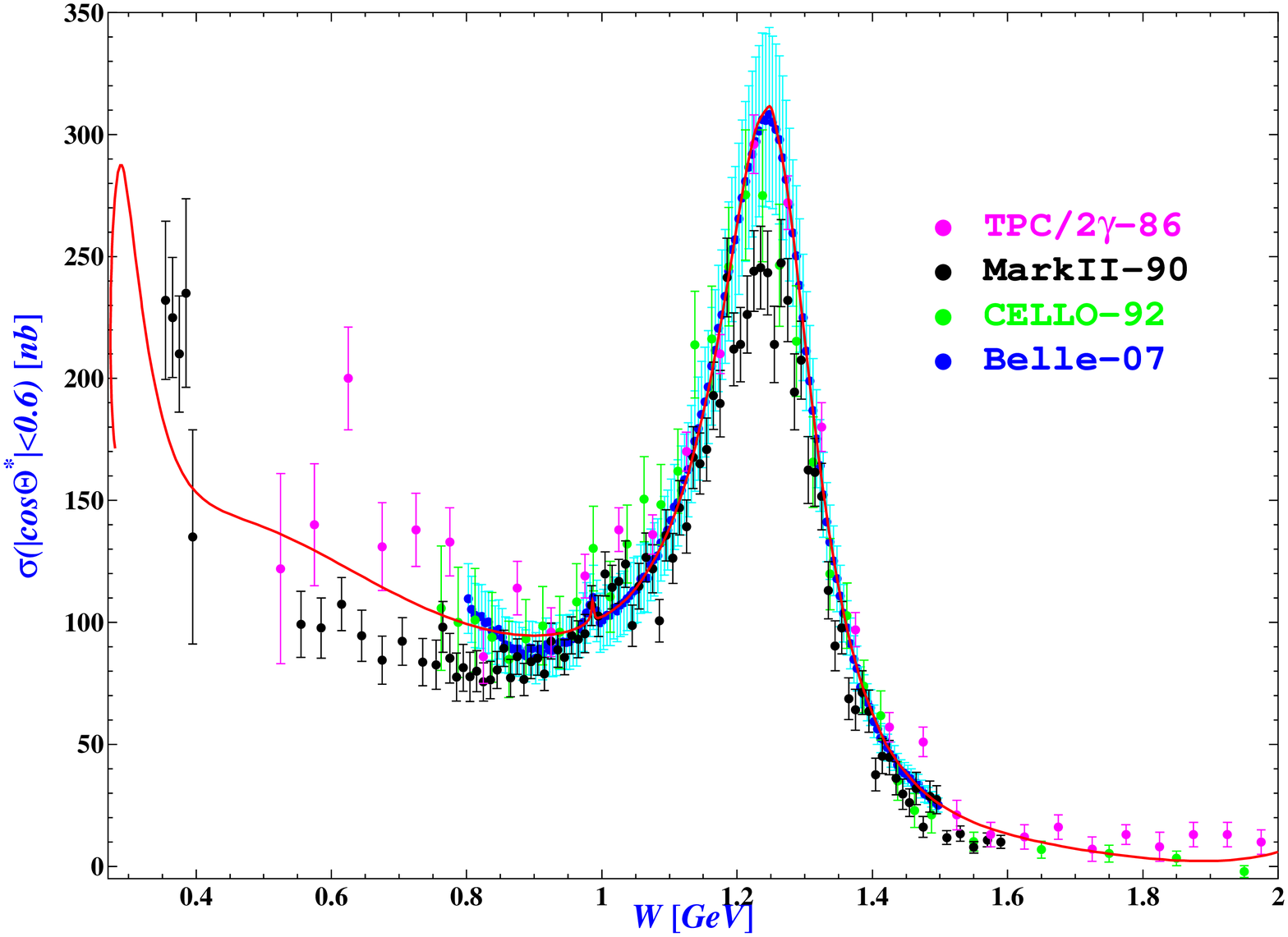}} 
\caption{
The cross section of the process $\gamma\gamma\to\pi^+\pi^-$ 
(with $|\cos\theta_{cm}|<0.6)$. Experimental data of TPC/2$\gamma$ \cite{tpc}, 
Mark II \cite{mark}, and CELLO \cite{cello} Collaborations are shown with 
statistical uncertainties only. Statistical uncertainties for the most of the 
Belle Collaboration data \cite{belle} are smaller then corresponding blue 
circles. Vertical light-blue error bars are systematic uncertainties 
for these data. The red line is our global fit result.}
\end{figure*}

Investigation of the $\gamma\gamma\to\pi^+\pi^-$ process was carried
out in the frameworks of different theoretical models
and, in particular, within dispersion relations 
(DR)~\cite{kal,don,bab,bij,garcia,hofer}.
Authors of most dispersion approaches restricted the partial
wave content by $s$ and $d$ waves only.  Moreover, they often used
additional assumptions, for example, to determine  subtraction
constants. The pion polarizabilities  
in a number of the works were obtained  from the analysis of the experimental
data in the region of the low energy ($ W<700$ MeV). 
The most of results for the charged pion polarizabilities
obtained in these works are close to the ChPT prediction~\cite{gasser2,burgi}.
On the other hand, the values
of the experimental cross section of the process $\gamma\gamma\to\pi^+\pi^-$ 
in this region are very ambiguous. Therefore, as it
has been shown in Refs.~\cite{don,fil3},
even changes of these values of the polarizabilities by more than 100\% are 
still compatible with the present error bars in the energy region considered.
More realistic values of the polarizabilities could 
be obtained analyzing the experimental data on $\gamma\gamma\to\pi^+\pi^-$       
in a wider energy region.

The processes $\gamma\gamma\to\pi^0\pi^0$ and $\gamma\gamma\to\pi^+\pi^-$ 
were analyzed in Refs.~\cite{fil3,fil1,fil2}
using DR with subtractions for the invariant amplitudes $M_{++}$ and
$M_{+-}$ without an expansion over partial waves. The subtraction constants
are uniquely determined in these works through the pion polarizabilities.
The values of polarizabilities have been found from the fit to the
experimental data of the processes $\gamma\gamma\to\pi^+\pi^-$ and 
$\gamma\gamma\to\pi^0\pi^0$ up to 2500 MeV and
2250 MeV, correspondently. As a result the following  values of
$\am_{\pi^{\pm}}=13.0^{+2.6}_{-1.9}$ and $\am_{\pi^0}=-1.6\pm 2.2$ have been
found in these works. In addition, for the first time there were obtained 
quadrupole polarizabilities for both charged and neutral pions.   

The new fit to the total cross section of the process
$\gamma\gamma\to\pi^+\pi^-$ at $|\cos\theta^{cm}_{\gamma\gamma}|<0.6$ in the 
frame of the DR~\cite{fil3} has been performed  with
the $\sigma$-meson considered as a pole on the second Riemann sheet.
The DR for the charged pions were saturated by the contributions of
the $\rho(770)$, $b_1(1235)$, $a_1(1270)$, and $a_2(1320)$ mesons in
the $s$ channel and $\sigma$, $f_0(980)$, $f_0^{\prime}(1500)$, $f_0(1710)$,
$f_0(2020)$, $f_2(1270)$, and $f_2(1565$ in the $t$ channel. 

As the two $K$ mesons give a big contribution to the decay width of the
$f_0(980)$ meson and the threshold of the reaction 
$\gamma\gamma\to K\overline{K}$ is very close to the mass of the 
$f_0(980)$ meson,
the Flatt{\'e} approximation~\cite{flatte} for $f_0(980)$ meson 
contribution was used.
Besides we took 
into account a nonresonance contribution of the $s$ waves with the isospin
$I=0$ and 2 using $\pi^+\pi^-$ loop diagrams.
The fit result  
using Eq.~(\ref{imm}) for $Im M_{++}^{\sigma}(t,s=\mu^2)$ with
the following parameters of the $\sigma$-meson: $M_{\sigma}=441$ MeV,
$\Gamma_{\sigma}=544$ MeV, $\ggg=1.298$ keV, $\gp=3.31$ GeV,       
is shown in Fig.~2. As a result of the fit we have obtained: 
$\am_{\pi^{\pm}}=10^{+2.9}_{-1.6}$ and $\ap_{\pi^{\pm}}=0.11^{+0.09}_{-0.02}$,
that agrees well within the errors with our previous fit~\cite{fil3} and
predictions of the Dispersion Sum Rules (DSR), see Table I in Ref.~\cite{fil2}.  

It should be noted that in the
region of $W\lesssim 500$ MeV the main contribution is given by the Born term,
the dipole and quadrupole polarizabilities, and the $\sigma$-meson. 
It would be very important to have new more accurate data 
in this energy region. 
                                                                                                   
\section{DSR and ChPT}

Here we discuss possible reason of the disagreement between DSR and ChPT
predictions for the charge pion polarizabilities.
DSR for the difference and sum of electric and magnetic pion polarizabilities
have been constructed in Refs.~\cite{fil3,fil1,fil2,bern}.
The main contribution to DSR for $\am_{\pi^{\pm}}$ is given by
$\sigma$-meson. However, this meson is taken into account only partially
in the ChPT calculations~\cite{gasser2}.

In the case of the $\pi^0$-meson,
the big contribution of the $\sigma$-meson to DSR is cancelled by the big 
contribution of the $\omega$-meson. 
The contribution of vector mesons in DSR can be written in the narrow width 
approximation as
\begin{equation}\label{vdsr}
Re M_{++}(s=\mu^2,t=0)=\frac{-4g_V^2M_V^2}{(M_V^2-\mu^2)}.
\end{equation}
In the case of ChPT, the authors of Ref.~\cite{gasser2} used
\begin{equation}\label{vchpt}
Re M_{++}(s=\mu^2,t=0)=\frac{-4g_V^2\mu^2}{(M_V^2-\mu^2)}.
\end{equation}
The absolute value of the amplitude (\ref{vchpt}) is smaller than (\ref{vdsr}) 
by a factor $M_V^2/\mu^2$. 
So, $\sigma$-meson is included in the ChPT calculations only partially and, 
according Eq. (\ref{vchpt}), the $\omega$-meson also gives a very small 
contribution. As a result, the predictions of DSR and ChPT for $\am_{\pi^0}$ are 
very close, see Table II in Ref.~\cite{fil2}.

\section{Summary and conclusions}

We have considered the main experimental works concerning charge
pion polarizabilities. 
The values of $\am_{\pi^{\pm}}$ obtained in the Serpukhov~\cite{serp}, 
Mainz~\cite{mami}, and LPI~\cite{lebed} experiments are at variance
with the ChPT predictions~\cite{gasser2}.
The result of the COMPASS Collaboration~\cite{comp} is in agreement with 
the ChPT calculations. However, this result is very model dependent.
It is necessary to correctly investigate the interference  between    
Coulomb and nuclear amplitudes and take into account the  
contribution of the $\sigma$-meson.

It should be noted that the most model independent result was obtained
in the Serpukhov experiment~\cite{serp}.

We have presented our new fit results for the $\gamma \gamma \to\pi^+ \pi^-$ reaction. 
The obtained value of $\am_{\pi^{\pm}}$ agrees well with
the DSR predictions and contradicts to ChPT where the $\sigma$-meson contribution
is taken into account only partially.  

In conclusion, further experimental and theoretical investigations are
needed to determine the true values of the pion polarizabilities.

The authors thank Th. Walcher and A.I. L'vov for useful discussions.



\end{document}